\documentclass[conference]{IEEEtran}
\IEEEoverridecommandlockouts

\usepackage{cite}
\usepackage{amsmath,amssymb,amsfonts}
\usepackage{graphicx}
\usepackage{textcomp}
\usepackage{xcolor}
\def\BibTeX{{\rm B\kern-.05em{\sc i\kern-.025em b}\kern-.08em
    T\kern-.1667em\lower.7ex\hbox{E}\kern-.125emX}}

\usepackage{todonotes}
\usepackage[inkscapearea=page]{svg}

\usepackage{mathtools}
\usepackage{hyperref}
\usepackage{amssymb}
\usepackage{bm}
\usepackage{graphicx}
\usepackage{float}
\usepackage{amsmath}
\usepackage{tikz}
\usepackage{bbm}
\usepackage{multirow}
\usetikzlibrary{positioning, backgrounds, fit, shapes.arrows}
\usetikzlibrary{decorations.markings}
\usepackage{amsthm}
\usepackage[table]{xcolor}
\usepackage{cuted}
\usepackage{algorithmicx}
\usepackage{algorithm}
\usepackage{algpseudocode}
\usepackage[caption=false,font=footnotesize]{subfig}
\usepackage{enumitem}
\usepackage{caption}
\usepackage{subcaption}

\tikzset{block/.style = {draw, fill=white, rectangle,
		minimum height=3em, minimum width=2cm},
	input/.style = {coordinate},
	output/.style = {coordinate},
	pinstyle/.style = {pin edge={to-,t,black}}
	radiation/.style={{decorate,decoration={expanding waves,angle=90,segment   length=4pt}}}
	
}
\usepackage{smartdiagram}

\tikzstyle{block} = [draw, rectangle, minimum height=2em, minimum width=2em]
\tikzstyle{sum} = [draw, circle,minimum width=0.1 cm]
\tikzstyle{input} = [coordinate]
\tikzstyle{output} = [coordinate]
\tikzstyle{dummy} = [coordinate]
\tikzstyle{pinstyle} = [pin edge={to-,thin,black}]
\usetikzlibrary{positioning, fit, arrows.meta}
\usetikzlibrary{positioning}
\usetikzlibrary{shapes,arrows}
\tikzstyle{frame_cyan} = [thick, draw=blue, solid,inner sep=0.3em]
\tikzstyle{frame_red} = [thick, draw=red, solid,inner sep=0.3em]
\tikzstyle{frame_green} = [thick, draw=green, solid,inner sep=0.3em]

\usepackage{tikz}
\usepackage{xcolor}
\definecolor{fc}{HTML}{1E90FF}
\definecolor{h}{HTML}{228B22}
\definecolor{bias}{HTML}{87CEFA}
\definecolor{noise}{HTML}{8B008B}
\definecolor{conv}{HTML}{FFA500}
\definecolor{pool}{HTML}{B22222}
\definecolor{up}{HTML}{B22222}
\definecolor{view}{HTML}{FFFFFF}
\definecolor{bn}{HTML}{FFD700}
\tikzset{fc/.style={black,draw=black,fill=fc,rectangle,minimum height=1cm}}
\tikzset{h/.style={black,draw=black,fill=h,rectangle,minimum height=1cm}}
\tikzset{bias/.style={black,draw=black,fill=bias,rectangle,minimum height=1cm}}
\tikzset{noise/.style={black,draw=black,fill=noise,rectangle,minimum height=1cm}}
\tikzset{conv/.style={black,draw=black,fill=conv,rectangle,minimum height=1cm}}
\tikzset{pool/.style={black,draw=black,fill=pool,rectangle,minimum height=1cm}}
\tikzset{up/.style={black,draw=black,fill=up,rectangle,minimum height=1cm}}
\tikzset{view/.style={black,draw=black,fill=view,rectangle,minimum height=1cm}}
\tikzset{bn/.style={black,draw=black,fill=bn,rectangle,minimum height=1cm}}
 
\usepackage{xspace}

\tikzstyle{dummy} = [coordinate]
\pgfkeys{/pgf/.cd,
  parallelepiped offset x/.initial=2mm,
  parallelepiped offset y/.initial=2mm
}
\pgfdeclareshape{parallelepiped}
{
  \inheritsavedanchors[from=rectangle] 
  \inheritanchorborder[from=rectangle]
  \inheritanchor[from=rectangle]{north}
  \inheritanchor[from=rectangle]{north west}
  \inheritanchor[from=rectangle]{north east}
  \inheritanchor[from=rectangle]{center}
  \inheritanchor[from=rectangle]{west}
  \inheritanchor[from=rectangle]{east}
  \inheritanchor[from=rectangle]{mid}
  \inheritanchor[from=rectangle]{mid west}
  \inheritanchor[from=rectangle]{mid east}
  \inheritanchor[from=rectangle]{base}
  \inheritanchor[from=rectangle]{base west}
  \inheritanchor[from=rectangle]{base east}
  \inheritanchor[from=rectangle]{south}
  \inheritanchor[from=rectangle]{south west}
  \inheritanchor[from=rectangle]{south east}
  \backgroundpath{
    \southwest \pgf@xa=\pgf@x \pgf@ya=\pgf@y
    \northeast \pgf@xb=\pgf@x \pgf@yb=\pgf@y
    \pgfmathsetlength\pgfutil@tempdima{\pgfkeysvalueof{/pgf/parallelepiped offset x}}
    \pgfmathsetlength\pgfutil@tempdimb{\pgfkeysvalueof{/pgf/parallelepiped offset y}}
    \def\ppd@offset{\pgfpoint{\pgfutil@tempdima}{\pgfutil@tempdimb}}
    \pgfpathmoveto{\pgfqpoint{\pgf@xa}{\pgf@ya}}
    \pgfpathlineto{\pgfqpoint{\pgf@xb}{\pgf@ya}}
    \pgfpathlineto{\pgfqpoint{\pgf@xb}{\pgf@yb}}
    \pgfpathlineto{\pgfqpoint{\pgf@xa}{\pgf@yb}}
    \pgfpathclose
    \pgfpathmoveto{\pgfqpoint{\pgf@xb}{\pgf@ya}}
    \pgfpathlineto{\pgfpointadd{\pgfpoint{\pgf@xb}{\pgf@ya}}{\ppd@offset}}
    \pgfpathlineto{\pgfpointadd{\pgfpoint{\pgf@xb}{\pgf@yb}}{\ppd@offset}}
    \pgfpathlineto{\pgfpointadd{\pgfpoint{\pgf@xa}{\pgf@yb}}{\ppd@offset}}
    \pgfpathlineto{\pgfqpoint{\pgf@xa}{\pgf@yb}}
    \pgfpathmoveto{\pgfqpoint{\pgf@xb}{\pgf@yb}}
    \pgfpathlineto{\pgfpointadd{\pgfpoint{\pgf@xb}{\pgf@yb}}{\ppd@offset}}
  }
}
\pgfdeclareshape{document}{
\inheritsavedanchors[from=rectangle] 
\inheritanchorborder[from=rectangle]
\inheritanchor[from=rectangle]{center}
\inheritanchor[from=rectangle]{north}
\inheritanchor[from=rectangle]{north east}
\inheritanchor[from=rectangle]{north west}
\inheritanchor[from=rectangle]{south}
\inheritanchor[from=rectangle]{south east}
\inheritanchor[from=rectangle]{south west}
\inheritanchor[from=rectangle]{west}
\inheritanchor[from=rectangle]{east}
\backgroundpath{%
\southwest \pgf@xa=\pgf@x \pgf@ya=\pgf@y
\northeast \pgf@xb=\pgf@x \pgf@yb=\pgf@y
\pgf@xc=\pgf@xb \advance\pgf@xc by-5pt 
\pgf@yc=\pgf@ya \advance\pgf@yc by5pt
\pgfpathmoveto{\pgfpoint{\pgf@xa}{\pgf@ya}}
\pgfpathlineto{\pgfpoint{\pgf@xa}{\pgf@yb}}
\pgfpathlineto{\pgfpoint{\pgf@xb}{\pgf@yb}}
\pgfpathlineto{\pgfpoint{\pgf@xb}{\pgf@yc}}
\pgfpathlineto{\pgfpoint{\pgf@xc}{\pgf@ya}}
\pgfpathclose
\pgfpathmoveto{\pgfpoint{\pgf@xc}{\pgf@ya}}
\pgfpathlineto{\pgfpoint{\pgf@xc}{\pgf@yc}}
\pgfpathlineto{\pgfpoint{\pgf@xb}{\pgf@yc}}
\pgfpathclose
}
}
\tikzstyle{block} = [draw, fill=white, rectangle, minimum height=3em, minimum width=6em]
    
\usetikzlibrary{backgrounds}
\usepackage{pifont}
\tikzstyle{startstop} = [rectangle, rounded corners, minimum width=0.8cm, minimum height=0.8cm,text centered, draw=black, fill=lime!30]
\usetikzlibrary{chains}
\usepackage{booktabs}
\usepackage{multirow}
\newcommand\MakeUppercaseGreek[1]{
  \begingroup
    \let\psi\Psi
    \let\omega\Omega
    \let\gamma\Gamma
    \MakeUppercase{#1}
  \endgroup}
\newcommand{\randomvec}[1]{\MakeUppercaseGreek{\bm{#1}}}
\newcommand{\vectorsym}[1]{\bm{#1}}
\newcommand{\expectation}[2]{\mathbb{E}_{#2}\left[#1\right]}

\newcommand{\brackets}[1]{\left(#1\right)}

\newcommand{\x}[0]{\vectorsym{x}}

\newcommand{\onsetbias}[0]{\mathcal{B}}
\newcommand{\onsetbiasnoise}[0]{\widetilde{\mathcal{B}}}

\begin{document}

\title{Low-Latency State Space Voice Activity Detection with Robust Onset Time Evaluation}

\author{
\IEEEauthorblockN{Elad Cohen}
\IEEEauthorblockA{
\textit{Arm Holdings}\\
Israel\\
elad.cohen2@arm.com}
\and
\IEEEauthorblockN{Arnon Netzer}
\IEEEauthorblockA{
\textit{Arm Holdings}\\
Israel\\
arnon.netzer@arm.com}
\and
\IEEEauthorblockN{Hai Victor Habi}
\IEEEauthorblockA{
\textit{Arm Holdings}\\
Israel\\
haivictor.habi@arm.com}
}

\maketitle

\begin{abstract}
Voice Activity Detection (VAD) systems are commonly evaluated using metrics such as the area under the receiver operating characteristic curve (AUROC), but these metrics do not account for temporal responsiveness. For low-latency applications, however, accurately measuring speech onset delay is essential. This is particularly challenging because onset latency evaluation is affected by noise and systematic misalignment in annotation timestamps. In this work, we introduce a probabilistic framework for evaluating VAD onset time under noisy temporal labels. We model annotated onset times as noisy observations of latent acoustic onsets and estimate the resulting discrepancy distribution from data. We show that this approach provides a more stable and robust estimate of algorithmic latency. In addition, we introduce S4VAD, the first state-space-model-based (SSM-based) VAD architecture. S4VAD supports efficient streaming inference while directly controlling the decay of past acoustic evidence, enabling fast and responsive speech-onset detection. We evaluate VAD onset latency across several low-latency architectures, including CNN, Transformer, and RNN models. Our proposed VAD achieves the lowest latency while maintaining a competitive AUROC.
\end{abstract}


\raggedbottom
\section{Introduction}

Voice Activity Detection (VAD) \cite{wang2025sincqdr,jia2021marblenet,ramirez2004efficient,kim2018voice,hughes2013recurrent} is a core component in speech processing pipelines, including speech recognition \cite{prabhavalkar2023survey,gulati2020conformer}, speaker diarization \cite{park2022review,anguera2012diarization}, speech enhancement \cite{wang2018supervised}, and real-time communication. VAD is usually evaluated with classification metrics such as accuracy, F1-score, equal error rate (EER), and area under the receiver operating characteristic curve (AUROC), but these metrics are largely insensitive to detection timing. In low-latency and streaming scenarios \cite{he2019streaming}, however, onset latency directly affects user experience and downstream system behavior. This sensitivity has become increasingly important with the rise of on-device and real-time audio user interfaces, where efficient, low-latency inference is critical for responsiveness and scalability. Despite its practical importance, algorithmic latency is much less standardized than classification accuracy in the VAD literature. It is typically defined as the time difference between an annotated speech onset and the first frame where the VAD output crosses a threshold or indicates speech \cite{ramirez2007statistical,tan2010integrated}. This definition implicitly assumes that onset annotations are temporally precise, an assumption that is often violated in practice.

\begin{figure}[t]
\centering
\includegraphics[width=0.95\linewidth]{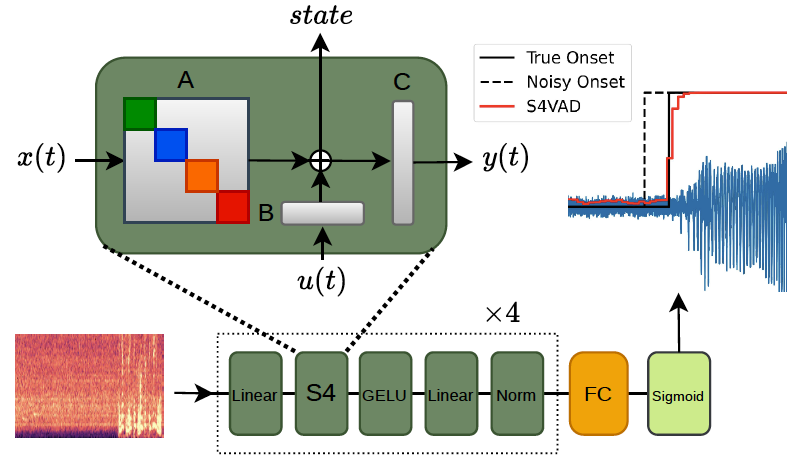}
\caption{\textbf{S4VAD.} Model Overview.}
\label{fig:s4vad}
\end{figure}

In this work, we demonstrate that onset annotations in widely used VAD datasets are inherently noisy, owing to ambiguous acoustic transitions and varying annotation conventions. In particular, after manually re-annotating a subset of these datasets for true onset times, we observe jitters on the order of tens to hundreds of milliseconds. As a result, naive latency measurements relative to annotated onset times can be unstable and misleading. Related areas such as speaker diarization and sound event detection \cite{mesaros2016metrics} address boundary uncertainty using collars or tolerance windows. While these reduce over-penalization of small temporal shifts, they mask boundary errors rather than provide explicit delay estimates, and therefore do not offer a principled solution for onset-latency evaluation under noisy labels.

To address this limitation, we propose an algorithmic onset-latency evaluation that explicitly accounts for annotation uncertainty. Rather than treating annotated onsets as exact timestamps, we model them as noisy observations of unknown true acoustic onsets. We estimate the annotation-noise statistics from noisy onset labels and reference VAD scores using Expectation–Maximization (EM) \cite{moon1996expectation}, and use the resulting mean offset to correct naive latency estimates. This produces robust and stable latency measurements under realistic annotation jitter. In addition, we introduce S4VAD, the first state-space model (SSM) \cite{gu2021combining} VAD, to the best of our knowledge. S4VAD is built on the S4 architecture \cite{gu2022parameterization}. It combines a linear state-space recurrence with efficient sequence modeling, controlling how quickly past acoustic evidence decays and providing a principled architectural mechanism for trading long-context robustness against fast onset responsiveness. We evaluate several low-latency VAD architectures, including CNN-, RNN-, and Transformer-based models, using the proposed framework. Our experiments show that S4VAD achieves the lowest onset latency while maintaining competitive AUROC.

Our main contributions are as follows:
\begin{itemize}
\item We propose an algorithmic onset-latency evaluation method for VAD models that explicitly accounts for annotation uncertainty, enabling more stable latency estimates.
\item We introduce S4VAD, the first SSM-based VAD. S4VAD enables efficient streaming inference while preserving fast local responsiveness for onset detection.
\item We present a systematic comparison of low-latency VAD models, including CNN, RNN, Transformer, and SSM-based models, with a focus on temporal responsiveness. Our proposed VAD achieves the lowest onset latency while maintaining a competitive AUROC.
\end{itemize}


\section{Background}

\subsection{Related Work}
Voice activity detection (VAD) has been studied using both classical signal-processing methods and modern neural architectures. Traditional approaches rely on features such as energy, zero-crossing rate, and statistical models \cite{sohn1999statistical, ramirez2004efficient}, while recent work employs CNNs, RNNs, and Transformers \cite{zhang2019end, jia2021marblenet, lavechin2020end, papadopoulos2021vadtransformer}. Although these methods achieve strong frame-level performance under metrics such as AUROC and EER, such metrics do not capture temporal responsiveness, which is crucial in streaming applications.

Latency-sensitive speech detection has mainly been studied for endpointing in streaming ASR. Prior work examined the trade-off between detection delay and false alarms \cite{arsikere2014endpointing, shannon2017endofquery}, while neural end-of-utterance detectors improved responsiveness in noisy conditions \cite{chang2017gridlstm}. However, these studies focus on end-of-speech detection and do not provide a general framework for onset evaluation under annotation uncertainty. Related domains address temporal uncertainty using tolerance regions. Sound event detection (SED) commonly applies onset/offset collars \cite{mesaros2016metrics, lafay2017dcase}, while diarization protocols use forgiveness regions \cite{anguera2012diarization}. These approaches reduce over-penalization but rely on heuristic windows rather than explicitly modeling annotation noise. Recent low-latency VAD work has considered streaming deployment and latency-aware evaluation. Semantic VAD \cite{shi2023semantic} reduces end-of-utterance delay using semantic information, while Personal VAD 2.0 \cite{ding2022personalvad2} targets resource-efficient personalized VAD for streaming ASR. Similarly, \cite{buddi2024pvad} explicitly evaluates detection latency in personalized VAD systems. However, personalized VAD additionally determines whether speech originates from an enrolled target speaker, and these works do not address onset evaluation under noisy temporal annotations. Speech onset detection can also be formulated as sequential change-point detection. Methods such as Bayesian online change-point detection \cite{adams2007bocd} and quickest change detection \cite{banerjee2014quickest, tartakovsky2015sequential} provide principled event-time estimation under uncertainty. Structured approaches, including CRFs and state-based models, have also been applied to VAD \cite{saito2010crfvad, fujimoto2007kalman}, but without handling noisy onset annotations.

\subsection{Onset Time Uncertainty}
Onset annotations are inherently imprecise due to human labeling variability \cite{yuan2013towards}, and recent studies report systematic onset deviations relative to human annotations \cite{gnanadesikan2021alignment, wu2023prosody}. AVA-Speech \cite{chaudhuri2018ava} provides annotations at 1~Hz resolution, leading to boundary uncertainty of hundreds of milliseconds. LibriSpeech \cite{panayotov2015librispeech} does not include manually verified word-level timestamps; boundaries are typically obtained using forced alignment tools such as Kaldi \cite{povey2011kaldi} or MFA \cite{mcauliffe2017mfa, rousso2024tradition}. Evaluations against manually aligned corpora such as TIMIT \cite{garofolo1993timit} and Buckeye \cite{pitt2005buckeye} show boundary deviations of tens of milliseconds \cite{mackenzie2020darla}. Thus, even relatively accurate alignments introduce non-negligible temporal uncertainty. This motivates evaluation methods that explicitly account for label uncertainty. We estimate onset time under noisy timestamp annotations by jointly modeling the true onset and annotation-noise distribution within a probabilistic evaluation framework.

\subsection{State Space Models (SSM)}
State-space models (SSMs) \cite{gu2021combining,gu2022efficiently,gu2022parameterization, gu2024mamba} describe sequential data through latent dynamical systems with explicit temporal memory. In continuous time, a linear SSM is defined as
\begin{equation}
\dot{x}(t) = A x(t) + B u(t), \qquad y(t) = C x(t),
\end{equation}
where $u(t)$ is the input signal, $x(t)\in\mathbb{C}^{N}$ is a latent state, and $y(t)$ is the output. Here, the parameters consist of the state matrix $A \in \mathbb{C}^{N\times N}$ and the input and output matrices $B \in \mathbb{C}^{N\times 1}$ and $C \in \mathbb{C}^{1\times N}$, respectively. State matrix $A$ governs the temporal dynamics of the state, while $B$ and $C$ control input coupling and readout. When the real parts of the eigenvalues of $A$ are negative, the system is stable and exhibits decaying memory exponentially.

For discrete-time sequence modeling, the continuous system is discretized with step size $\Delta$, yielding
\begin{equation}
x_{t+1} = \bar{A} x_t + \bar{B} u_t, \qquad y_t = C x_t,
\end{equation}
where $\bar{A} = \exp(\Delta A)$ and $\bar{B} = (\Delta A)^{-1} (\exp(\Delta A) - I) \Delta B$. This formulation \cite{gupta2022diagonal} makes temporal memory explicit: past inputs influence future outputs through the latent state, with a decay rate determined by $\bar{A}$.

\newcommand{\tstar}[0]{t^*}
\newcommand{\that}[0]{\widehat{t}}
\section{Onset Time Evaluation}
Here, we present an onset time evaluation method for VAD models that explicitly accounts for annotation noise. Specifically, let's denote $\tstar$ as the true onset time, $t$ its observed (noisy) annotation with annotator noise $\delta\sim \pi(\delta)$, and $\x$ represents the audio segment around the annotated onset. Then, the value of $t$ can be written as:
\begin{align}\label{eq:noisy_label}
t=t^{*}+\delta.
\end{align}
Given a target VAD, our objective is to derive an onset evaluation criterion using a noisy label dataset $\mathcal{D}=\{\x_{i},t_{i}\}_{i=1}^N$ consisting of $N$ i.i.d. samples. To derive such an evaluation, we first assume that $\pi(\delta)$ is a bounded noise distribution such that $\pi(\delta)=0$ for $\delta\not\in [\delta_L,\delta_R]$. Using the assumption and target VAD output scores $s(t;\x)$, we define the estimated onset time as: 
\begin{align}
\that(\x) = \min_{\tilde{t}\in\mathcal{T}_{\x}}\{\tilde{t} : \forall u \in [\tilde{t},\, \tilde{t}+T],\; s(u;\x) \ge \tau\},
\end{align}
where $\mathcal{T}_{\x}$ is the time samples of  $\x$, $\tau$ is a threshold, and $T>0$ is a time interval. In this work, we define the algorithmic latency as \emph{onset bias}, i.e., the average temporal delay between a VAD decision and the true speech onset. The expected onset bias is then defined by: 
\begin{align}\label{eq:onset_metric}
&\mathcal{B}=\expectation{{\that\brackets{\x}-\tstar}}{}=\underbrace{\expectation{\epsilon}{}}_{\widetilde{\mathcal{B}}}+{\color{red}\underbrace{\expectation{\delta}{}}_{\mu_{\delta}}},
\end{align}
where $\epsilon\triangleq\that\brackets{\x}-t$, $\mu_\delta$ denotes the mean error of the annotation noise, and $\widetilde{\mathcal{B}}$ is the naive onset bias without the annotation error correction. We define the onset latency as the sum of the corrected onset bias $\mathcal{B}$ and the hardware inference latency, since a streaming system can emit a speech decision only after both the model’s decision delay and the computation time have elapsed. Our objective is to estimate the average onset bias $\mathcal{B}$ at the dataset level in the presence of annotation noise. To estimate the onset bias $\mathcal{B}$ from a dataset $\mathcal{D}$, we propose the following procedure. First, we learn a single global noise model for the entire dataset using Expectation-Maximization (EM) in conjunction with a reference VAD model. Then, we use the dataset $\mathcal{D}$ together with the noise model to evaluate $\mathcal{B}$.

\subsection{Annotator Noise Distribution Estimation}\label{sec::noise}
Annotation timestamps are often noisy and systematically misaligned with true acoustic onsets, making direct onset measurements unreliable.  We therefore estimate the noise distribution from the data to capture this uncertainty and obtain a robust estimate of the onset bias $\mathcal{B}$. Specifically, we model the annotation noise and estimate its distribution using the EM algorithm \cite{moon1996expectation}.  

\noindent\textbf{Expectation Step (E-step):} Here, we derive the E-step for the EM. Specifically, let us define the expected complete-data log-likelihood: 
\begin{align}\label{eq:q}
    Q\brackets{\pi|\pi^{(k)}}&=\expectation{\log P\brackets{\mathcal{D},\randomvec{\delta}|\pi}}{\randomvec{\delta}\sim P\brackets{\cdot|\mathcal{D},\pi^{(k)}}}\\
    &=\sum_{i=1}^N\expectation{\log P\brackets{\x_i,t_i,\delta|\pi}}{\delta\sim P\brackets{\cdot|\x_i,t_i,\pi^{(k)}}}.\nonumber
\end{align}
To obtain the E-step, we need the posterior probability of the annotation error and the likelihood term. For the posterior probability, we apply Bayes’ rule, which gives:
\begin{equation}\label{eq:bayes_post}
    q\brackets{\delta}\triangleq P\brackets{\delta|\x,t,\pi}=\frac{P\brackets{t,\delta|\x,\pi}}{P\brackets{t|\x,\pi}}.
\end{equation}
Using the chain rule of probability, we have $P\brackets{t,\delta|\x,\pi}=P\brackets{t|\delta,\x}P\brackets{\delta|x,\pi}$. Assuming that $P\brackets{\delta \mid x,\pi} = \pi\brackets{\delta}$, i.e., the annotator error is independent of signal $\x$, we approximate $P\brackets{t|\delta,\x}$ by the following hazard-style approximation which yields:
\begin{align}\label{hazard_aprox}
P\brackets{t|\delta,\x}&\approx\psi(t-\delta;\x)\quad\text{where}\nonumber \\
\psi(t;\x)&\triangleq \frac{1}{Z}r(t;\x)\prod_{u<t}(1-r(u;\x)),
\end{align}
$r(t;\x)\in[0,1]$ is the activity score of a well-trained reference VAD and $Z$ is the normalization constant. Combining \eqref{hazard_aprox} and $P\brackets{\delta \mid x,\pi} = \pi\brackets{\delta}$ results in
\begin{equation}\label{eq:combine_prob}
    P\brackets{t,\delta|\x,\pi}=\pi(\delta)\, \psi(t - \delta;\x).
\end{equation}
Plugging \eqref{eq:combine_prob} into \eqref{eq:bayes_post}, we have the posterior probability of annotator error $\delta$ for sample $i$:
\begin{equation}\label{eq:q_n}
    q_i\brackets{\delta}=\frac{\pi(\delta)\, \psi(t_i - \delta;\x_i)}{\sum_{\delta'=\delta_L}^{\delta_R}\pi(\delta')\, \psi(t_i - \delta';\x_i)}.
\end{equation}
Next, we decompose the objective in \eqref{eq:q} as: 
\begin{align}
    \log P\brackets{\x_i,t_i,\delta|\pi}=\log P\brackets{\x_i}+\log P\brackets{t_i,\delta|\x_i, \pi}.
\end{align}
As the probability of $\x_i$ being independent of the annotation and by using \eqref{eq:combine_prob} we obtain:
\begin{align}\label{eq:q_final}
    Q\brackets{\pi|\pi^{(k)}}&=\sum_{i=1}^N\expectation{\log \pi\brackets{\delta}\psi\brackets{t_i-\delta;\x_i}}{\delta\sim P\brackets{\cdot|\x_i,t_i,\pi^{(k)}}}\nonumber\\
    &=\sum_{i=1}^N\sum_{\delta=\delta_L}^{\delta_R}q_i\brackets{\delta}\log \pi\brackets{\delta}.
\end{align}
In the second step of \eqref{eq:q_final}, we use the fact that $\psi$ is independent of $\pi$ and apply \eqref{eq:q_n}. \\
\noindent\textbf{Maximization Step (M-step):} In the M-step, we maximize the objective in \eqref{eq:q_final} while ensuring that $\pi$ is a valid distribution, by defining the constrained optimization problem:
\begin{align}\label{opt_prob}
\pi(\delta)=\arg\max_{\pi}
&\sum_{i=1}^{N}
\sum_{\delta=\delta_L}^{\delta_R}
q_i(\delta)\log \pi(\delta),\\
\text{s.t}\quad&\sum_{\delta=\delta_L}^{\delta_R}\pi(\delta)=1.\nonumber
\end{align}
Solving the optimization problem~\eqref{opt_prob} yields:
\begin{align}\label{pi_est}
\pi(\delta)=\frac{1}{N} \sum_{i=1}^{N} q_i(\delta).
\end{align}
Finally, we initialize $\pi$ to be a uniform distribution by
$\pi^{(0)}(\delta) = \frac{1}{\delta_R - \delta_L + 1}$, and then alternate between the E-step and M-step for $K$ iterations.

The estimated noise distribution $\pi(\delta)$ can capture both annotation noise and the systematic bias of the reference VAD, which are not identifiable separately without additional supervision. Consequently, we use a high-quality, well-trained reference VAD model, and therefore expect its systematic bias to be relatively small compared to annotation uncertainty. As a result, we interpret the estimated shift primarily as reflecting annotation noise, while acknowledging that residual reference-dependent bias may remain.

\begin{figure}[t]
    \centering
    \subfloat[LibriSpeech]{%
        \includegraphics[width=0.23\textwidth]{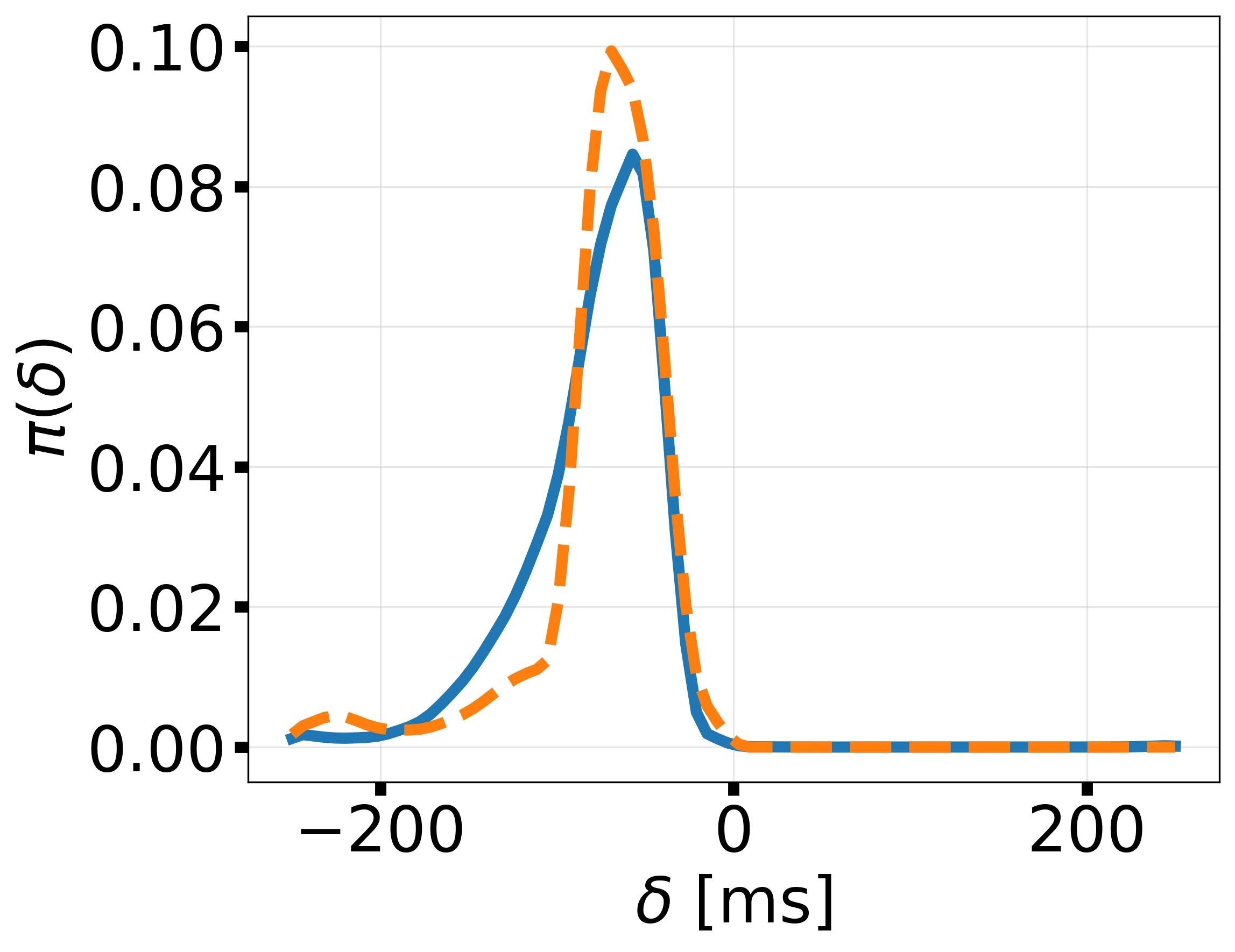}%
        \label{fig:onset_acam}
    }
    \hfill
    \subfloat[AVA-Speech]{%
        \includegraphics[width=0.23\textwidth]{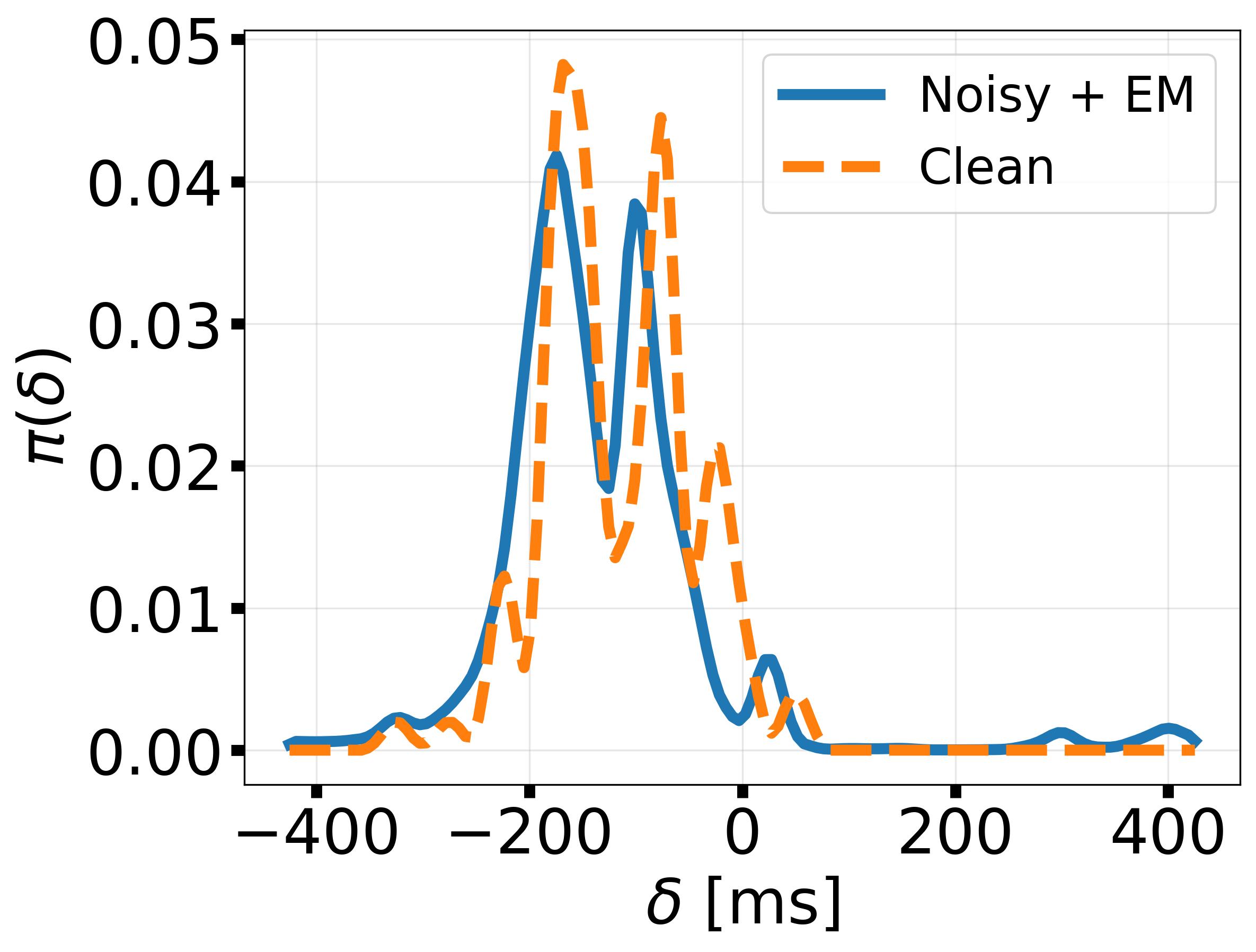}%
        \label{fig:onset_ava}
    }
    \caption{\textbf{Annotation noise distribution.} Estimated from a small subset of clean and noisy onset labels. As shown, outputs from our EM algorithm closely match the shape obtained from the clean subset. The negative values of $\delta$ indicate a systematic early-labeling bias, whereby annotators tend to place onset labels before the true onset. Results are reported for (a) LibriSpeech and (b) AVA-Speech.}
    \label{fig:pi}
\end{figure}

\subsection{Annotation Noise Analysis}
To mitigate bias from the reference VAD, we employ Silero VAD \cite{silero}, a lightweight, high-quality neural VAD widely used in real-time audio applications. To verify that its bias is substantially smaller than the annotation noise, we obtain \emph{clean onset labels}. Specifically, we collect 100 noisy onset instances from each dataset in Subsection~\ref{sec:imp} and manually identify their true onset times, $t^*$. We then compare the resulting annotation-noise histogram with our estimated $\pi(\delta)$. We run the EM algorithm for $K=5$ iterations, using $\delta_L=-250~\mathrm{ms}$ and $\delta_R=250~\mathrm{ms}$ for LibriSpeech, and $\delta_L=-500~\mathrm{ms}$ and $\delta_R=500~\mathrm{ms}$ for AVA-Speech, reflecting their different uncertainty levels. The estimated annotation-noise distributions are shown in Figure~\ref{fig:pi} and closely match those obtained from the manually annotated onsets, with mean errors of 5~ms and 13~ms for LibriSpeech and AVA-Speech, respectively. These results empirically support the assumptions in Subsection~\ref{sec::noise}. Moreover, $\pi(\delta)$ is sharply concentrated around its mean, suggesting that annotation noise is dominated by dataset-level bias rather than sample-specific variability. Thus, a constant dataset-level correction $\mu_\delta$ is preferable to per-sample adjustments that may introduce additional variance. 

Note that the reference VAD alone is insufficient: directly comparing a target VAD to the reference VAD would measure only relative latency and would inherit sample-level reference errors. The noisy label, despite its temporal bias, anchors the evaluation to the true onset event. Our method uses the noisy annotation to define a local candidate window and uses the reference VAD only within this window to infer the true clean onset. Within the EM procedure, the reference VAD acts as a weak local probabilistic observer, enabling the dataset-level annotation shift to be estimated from many noisy examples rather than from a small manually corrected subset. By taking the expectation of $\pi(\delta)$ on the full datasets, we obtain dataset-level shifts of $\mu_\delta =-70\,\mathrm{ms}$ and $\mu_\delta =-136\,\mathrm{ms}$ for LibriSpeech and AVA-Speech, respectively. These results indicate that the reference-VAD mean error is much smaller than the annotation shift being corrected.

\begin{table*}[t]
\centering
\resizebox{1.0\textwidth}{!}{
\begin{tabular}{c|c|ccc|ccc}
\toprule
\rowcolor[HTML]{FFCE93}
\textbf{Model}
& \textbf{HW Latency}
& \multicolumn{3}{c|}{\textbf{LibriSpeech}}
& \multicolumn{3}{c}{\textbf{AVA-Speech}} \\
\rowcolor[HTML]{FFCE93}
& [ms]
& AUROC
& Onset Bias [ms]
& Onset Latency [ms]
& AUROC
& Onset Bias [ms]
& Onset Latency [ms] \\
\midrule
MarbleNet \cite{jia2021marblenet} & 1.3 & 0.974 & 54 & 55.4 & 0.866 & 159 & 160.3 \\
TrVAD \cite{zhao2022efficient} & 2.7 & 0.950 & 50 & 52.7 & 0.851 & 122 & 124.7  \\
ResectNet \cite{kopuklu2022resectnet} & 0.6 & 0.980 & 51 &  51.6 & 0.823 &  237 & 237.6  \\ \midrule
    \textbf{S4VAD} (Ours) & \textbf{0.3} & 0.972 & \textbf{39} & \textbf{39.3} & 0.846 & \textbf{114} & \textbf{114.3}  \\
\bottomrule
\end{tabular}}
\caption{\textbf{Onset Latency.} Comparison of VAD latencies across different models for target FPR of 3\%.}
\label{table:onset}
\end{table*}

\section{SSM-based VAD}
Here, we present S4VAD, a VAD model based on a diagonal-structured state-space architecture (S4D). Structured State-Space models (S4) \cite{gu2021combining} constrain the state transition matrix $A$ to have a diagonal or diagonal-plus-low-rank structure, enabling efficient modeling of both short-term and long-term dependencies in sequential data. In this work, we use S4D variant \cite{gu2022parameterization} where $A$ is a diagonal matrix with complex-valued entries,
\begin{equation}
A = \operatorname{diag}(\lambda_0, \dots, \lambda_{N-1}), \qquad \lambda_n=-0.5+j\pi n,
\end{equation}
and the state update can be computed independently for each dimension. In addition, the input and output matrices are also constrained to be diagonal, such that each channel maintains an independent latent state driven by its corresponding input dimension. This design enables efficient computation and avoids cross-channel mixing within the state dynamics. The discretization step is sampled once per channel from the log uniform distribution:
\begin{equation}\label{eq:log_dist}
log(\Delta)\sim\mathcal{U}(log (dt_{min}), log (dt_{max})),
\end{equation}
where $dt_{min}$ and $dt_{max}$ are the minimum and maximum time step scales, respectively. $\Delta$ controls the hidden-state memory timescale: larger values make past information decay faster and yield more local, responsive dynamics, while smaller values preserve longer context but may smear speech evidence into silence. S4VAD operates on frame-level acoustic features and predicts a speech probability for each frame. The discrete-time form of S4VAD has an equivalent convolutional form with a kernel length of $L$:
\begin{equation}
y_t = \sum_{n=0}^{L-1} K_n u_{t-n},
\end{equation}
where the convolution kernel $K$ is real-valued and analytically derived from the SSM parameters as $K_n = 2 \Re(C \bar{A}^n \bar{B})$. This representation enables efficient parallel training using one-dimensional convolutions, while remaining exactly equivalent to the recurrent state update. At inference time, the model can be executed in streaming mode by updating the latent state sequentially using the recurrence in Eq.~(2). As a result, S4VAD combines efficient convolutional training with explicit state-based inference. The temporal behavior of the model is governed by interpretable time constants determined by the eigenvalues of $A$ and the discretization step $\Delta$, making S4VAD particularly well suited for streaming speech tasks such as voice activity detection.

\begin{table}[t]
\centering
\label{tab:latency_stats}
\resizebox{0.48\textwidth}{!}{
\begin{tabular}{lcccc}
\toprule
\rowcolor[HTML]{FFCE93}
\textbf{Model} & \textbf{Std} [ms] & \textbf{P50} [ms] & \textbf{P90} [ms] & \textbf{P95} [ms] \\
\midrule
MarbleNet \cite{jia2021marblenet}  & 87 & 10 & 150 & 210 \\
TrVAD \cite{zhao2022efficient}  & 73 & 10 & 110 & 160 \\
ResectNet \cite{kopuklu2022resectnet} & 74 & 10 & 100 & 170 \\
\midrule
\textbf{S4VAD} (Ours) & 66 & 10 & 90  & 130 \\
\bottomrule
\end{tabular}}
\caption{Onset latency confidence metrics for LibriSpeech.}
\label{table:conf}
\end{table}

\noindent\textbf{Architecture Details.}\indent A diagram of S4VAD is shown in Figure~\ref{fig:s4vad}. It consists of four S4 blocks with hidden dimension $d_{\text{model}}=64$. Each block projects the input to the model dimension, applies an S4 layer to capture long-range temporal dependencies, and then passes the output through GELU, a linear layer, and dropout. A residual connection and layer normalization are applied within each block. The S4 layer builds a convolution kernel of $L=201$ from continuous-time state space parameters, where the discretization step is sampled once per channel from~\eqref{eq:log_dist} with $dt_{min}=0.6$ and $dt_{max}=1.0$. The state dimension is $N=64$, implemented as $N/2$ complex conjugate pairs. For causal inference, asymmetric padding ensures that predictions depend only on past frames. After the S4 stack, a final layer normalization and a linear classification head produce one logit per frame. The model supports both offline convolutional inference and streaming inference, where the convolution is replaced by a recurrent state update for frame-by-frame processing with constant memory and compute.

\section{Experimental Results}
This section describes the experimental setup, implementation details, and results. The implementation details are provided in Subsection~\ref{sec:imp}. A comparison of several low-latency VAD models is described in Subsection~\ref{sec:comp}. Finally, an ablation study is presented in Subsection~\ref{sec:ablations}.

\subsection{Implementation Details}\label{sec:imp}
In this work, we use clean speech from LibriSpeech~\cite{panayotov2015librispeech} combined with noise from DNS Challenge~\cite{dubey2023icassp}. The DNS dataset comprises a diverse range of noise types, including real-world recordings (e.g., traffic, crowd noise, household appliances, and office environments) and synthetic noise. The noise samples were divided into training, validation, and test sets with an 8:1:1 ratio, while the original LibriSpeech splits were preserved; we used the 100-hour training subset. A representative VAD dataset was synthesized by concatenating background-noise segments with noisy speech segments. This created clear onset transitions, moving from noise to speech. Frame-level labels were assigned by majority voting, where a frame was labeled as speech if most of its samples corresponded to speech. In addition, we use the AVA-Speech~\cite{chaudhuri2018ava} dataset. AVA-Speech annotates real-world speech activity from YouTube movie clips and contains audio segments and their speech/noise labels. Onset samples are extracted by selecting a 1-second window around the noisy onset label, ensuring it contains only a single transition. Samples without a sustained threshold crossing are treated as missed detections (below 1\% for all models) and excluded from the onset-latency computation. The number of onset samples used is about 6,000 and 3,500 for LibriSpeech and AVA-Speech, respectively. The pre-processing consists of raw audio signals sampled at 16~kHz, which are then converted into log-Mel spectrograms. Specifically, each waveform was segmented into 32~ms frames with a 10~ms hop size and transformed into 64-dimensional log-Mel features.

\noindent\textbf{Training.}\indent Training was performed on the training split of the synthetic mixtures of LibriSpeech. Noisy mixtures were generated uniformly across signal-to-noise ratio (SNR) levels from $-5$~dB to $20$~dB. Additional white noise augmentation, with amplitudes ranging from $-90$~dB to $-46$~dB, was applied to improve robustness. Models were trained using binary cross-entropy loss and optimized with AdamW~\cite{loshchilov2017fixing} for 100 epochs with a batch size of 512. A cosine annealing learning rate schedule~\cite{loshchilov2016sgdr} was employed, with learning rates ranging from $10^{-3}$ to $5\times10^{-5}$, and model selection was based on validation AUROC.

\noindent\textbf{Evaluation.}\indent Evaluation was performed on the test split of the synthetic mixtures of LibriSpeech. Here, we mix speech with noise at SNR levels between $-10$~dB and $10$~dB, at $5$~dB steps. In addition, we evaluate the performance on AVA-Speech~\cite{chaudhuri2018ava}. The onset bias was calculated using the proposed method with $T = 50~\text{ms}$, to ensure that a detected rise corresponds to sustained speech activity rather than a transient spike. The threshold $\tau$ for each VAD is chosen such that it achieves a target False Positive Rate (FPR) on the validation set. For hardware inference latency, all models were lowered to the XNNPACK backend and evaluated on an Apple M4 Max 64GB as the CPU inference platform.

\subsection{Low-Latency VADs Comparison}\label{sec:comp}
We evaluate four low-latency VAD models: MarbleNet \cite{jia2021marblenet} (CNN-based), TrVAD \cite{zhao2022efficient} (Transformer-based), ResectNet \cite{kopuklu2022resectnet} (RNN-based), and our S4VAD (SSM-based). All of these models are reproduced and trained using the same preprocessing and operate in causal mode. During inference, each model produces frame-level speech posterior probabilities in a streaming mode at a hop size of 10 ms. Table~\ref{table:onset} reports AUROC, HW latency, onset bias, and onset latency under target FPRs of $3\%$. In addition, Table~\ref{table:conf} reports onset latency statistics, including the standard deviation and P50, P90, and P95 percentiles. As shown, S4VAD achieves the lowest onset latency while remaining competitive in AUROC. MarbleNet, as a CNN-based model, aggregates evidence over a fixed temporal window, which can delay onset decisions until sufficient post-onset information is available. TrVAD benefits from attention-based temporal modeling but incurs higher computational cost, reflected in its increased hardware latency. ResectNet is naturally stateful, updating its recurrent state as new frames arrive rather than recomputing a full context window. This may explain its low latency on LibriSpeech, while its higher latency on AVA-Speech suggests reduced robustness under challenging acoustic conditions. In contrast, S4VAD combines state-space sequence modeling with efficient streaming inference. Its compact recurrent state captures long-range dependencies while preserving fast local dynamics. Consequently, S4VAD provides low onset latency with AUROC comparable to the strongest baselines, making it well-suited for real-time VAD applications requiring early and accurate speech detection.

\begin{figure}[t]
\centering
\includegraphics[width=0.9\linewidth]{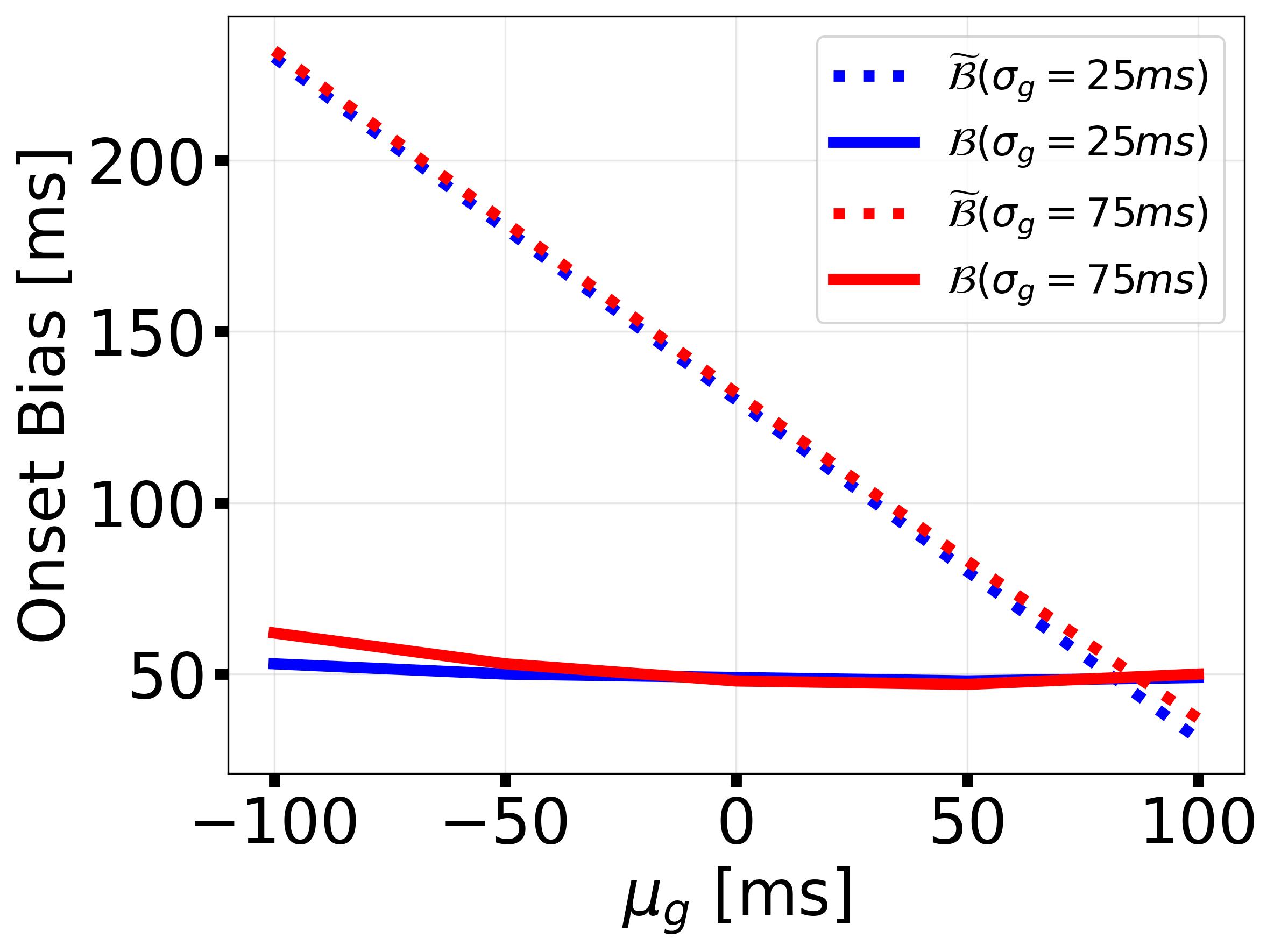}
\caption{\textbf{Naive and our Onset Bias.} Comparison between the naive onset bias $\onsetbiasnoise$ and our $\onsetbias$. Here, an additional Gaussian noise label $g\sim\mathcal{N}(\mu_g,\sigma_g^2)$ is introduced with mean $\mu_g$ and variance $\sigma^2_g$. As shown, our method remains stable and effectively tracks the noise label shift, whereas the naive method is highly affected.}
\label{fig:stable_onset}
\end{figure}

\begin{figure}[t]
\centering
\includegraphics[width=0.9\linewidth]{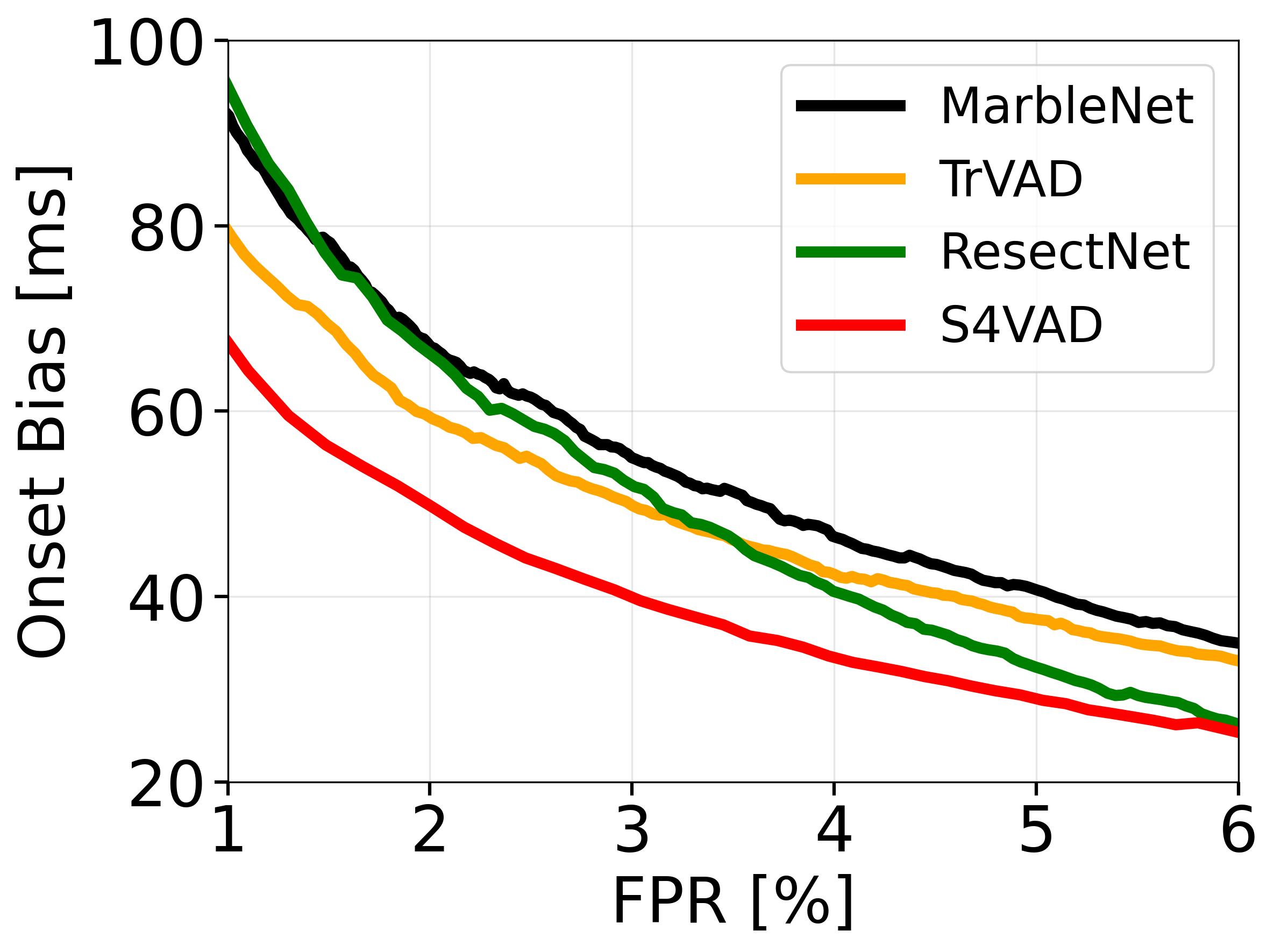}
\caption{\textbf{Onset Bias vs. FPR.} Evaluated on LibriSpeech.}
\label{fig:onset_ds_bias_fpr}
\end{figure}

\begin{table}[]
\centering
\begin{tabular}{c|cc|cc}
\toprule
\rowcolor[HTML]{FFCE93} 
\cellcolor[HTML]{FFCE93} & \multicolumn{2}{c}{\cellcolor[HTML]{FFCE93}\textbf{LibriSpeech}} & \multicolumn{2}{c}{\cellcolor[HTML]{FFCE93}\textbf{AVA-Speech}} \\ \cline{2-5} 
\rowcolor[HTML]{FFCE93} 
\multirow{-2}{*}{\cellcolor[HTML]{FFCE93}\textbf{$dt_{min}$}} & \multicolumn{1}{l}{\cellcolor[HTML]{FFCE93}AUROC} & \multicolumn{1}{l}{\cellcolor[HTML]{FFCE93}Onset Bias {[}ms{]}} & \multicolumn{1}{l}{\cellcolor[HTML]{FFCE93}AUROC} & \multicolumn{1}{l}{\cellcolor[HTML]{FFCE93}Onset Bias {[}ms{]}} \\ \midrule
0.1 & 0.979 & 65 & 0.856 & 156 \\
0.3 & 0.976 & 55 & 0.841 & 139 \\
0.6 & 0.972 & 39 & 0.846 & 114 \\
0.9 & 0.968 & 44 & 0.826 & 122 \\ \bottomrule
\end{tabular}
\caption{\textbf{Discretization step ablation.}}
\label{table:ablation}
\end{table}

\subsection{Ablations}\label{sec:ablations}

\noindent\textbf{Naive and our Onset Bias.} \indent In Figure~\ref{fig:stable_onset}, we compare the proposed onset bias evaluation with a naive implementation under controlled annotation noise. Specifically, we perturb the original onset time labels by adding Gaussian noise, $t_{\text{new}} = t + g$, where $g \sim \mathcal{N}(\mu_g, \sigma^2_g)$. We then apply the naive and our method to the perturbed labels and evaluate the resulting onset bias. The results show that the naive onset latency evaluation is strongly affected by annotation noise, whereas the proposed method adapts to the underlying noise distribution and yields more stable onset latency estimates. The evaluation was performed using MarbleNet for a 3\% FPR target, on the LibriSpeech test set.

\noindent\textbf{Onset Bias vs. FPR.} \indent Figure~\ref{fig:onset_ds_bias_fpr} illustrates the trade-off between onset bias and FPR. We evaluate onset bias across several FPR operating points. Lower FPR values correspond to more conservative detection, resulting in larger onset delays, whereas higher FPR values allow earlier detection. As shown, S4VAD exhibits the lowest onset bias across all FPRs. 

\noindent\textbf{S4VAD Ablation} \indent 
S4VAD provides an architectural control knob for the AUROC and onset bias tradeoff through the discretization timescale $\Delta$. As shown in Table~\ref{table:ablation}, varying $dt_{\min}$ changes the memory timescale and produces a systematic tradeoff between AUROC and onset bias. This behavior is specific to the SSM parameterization: larger $\Delta$ values encourage faster decay and more local responsiveness, while smaller values preserve longer context.
The results show that both overly short and overly long memory timescales can degrade onset responsiveness. Here, a value of $dt_{\min}=0.6$ provides the lowest onset bias while maintaining a competitive AUROC.

\section{Conclusions}
In this work, we introduced a probabilistic method for onset time estimation in Voice Activity Detection under noisy onset annotations. By modeling the annotated onset as a noisy realization of the true onset time, we formulated onset time evaluation as a statistical problem and derived corrected onset bias measurements. In addition, we introduce S4VAD, a state-space VAD architecture that preserves long-range temporal context while maintaining fast local dynamics for responsive streaming speech detection. Our experiments show that S4VAD achieves the fastest onset latency while maintaining a competitive AUROC.


\newpage
\bibliographystyle{IEEEtran}
\bibliography{IEEEexample}

\vspace{12pt}
\color{red}

\end{document}